\documentclass{vgtc}                          

\ifpdf
  \pdfoutput=1\relax                   
  \usepackage{graphicx}                
  \DeclareGraphicsExtensions{.pdf,.png,.jpg,.jpeg} 
\else
  \usepackage{graphicx}                
  \DeclareGraphicsExtensions{.eps}     
\fi%

\graphicspath{{figures/}{pictures/}{images/}{./}} 

\usepackage{microtype}                 
\PassOptionsToPackage{warn}{textcomp}  
\usepackage{textcomp}                  
\usepackage{mathptmx}                  
\usepackage{times}                     
\usepackage{cite}                      
\usepackage{tabu}                      
\usepackage{booktabs}                  
\usepackage{amsmath}
\usepackage{makecell}
\usepackage{multirow}

\onlineid{0}

\vgtccategory{Research}

\vgtcinsertpkg

\title{Dynamic Scene Adjustment for Player Engagement in VR Game}

\author{Zhitao Liu\thanks{e-mail: zl425uestc@gmail.com}\\ %
\and Yi Li\thanks{e-mail: elieli0925@qq.com}\\ %
\and Ning Xie\thanks{Corresponding author, e-mail: seanxiening@gmail.com}\\ %
\and Youteng Fan\thanks{e-mail: fyouteng@163.com}\\ %
\and Haolan Tang\thanks{e-mail: tanghl1203@163.com}\\ %
\and Wei Zhang\thanks{e-mail: 37058836@qq.com}\\ %
     \scriptsize AVIC Chengdu Aircraft Design $\&$ Research Institute, Chengdu, China
     }
\affiliation{\scriptsize Center for Future Media, the School of Computer Science and Engineering, UESTC, Chengdu, China}

\teaser{
 \includegraphics[width=\textwidth]{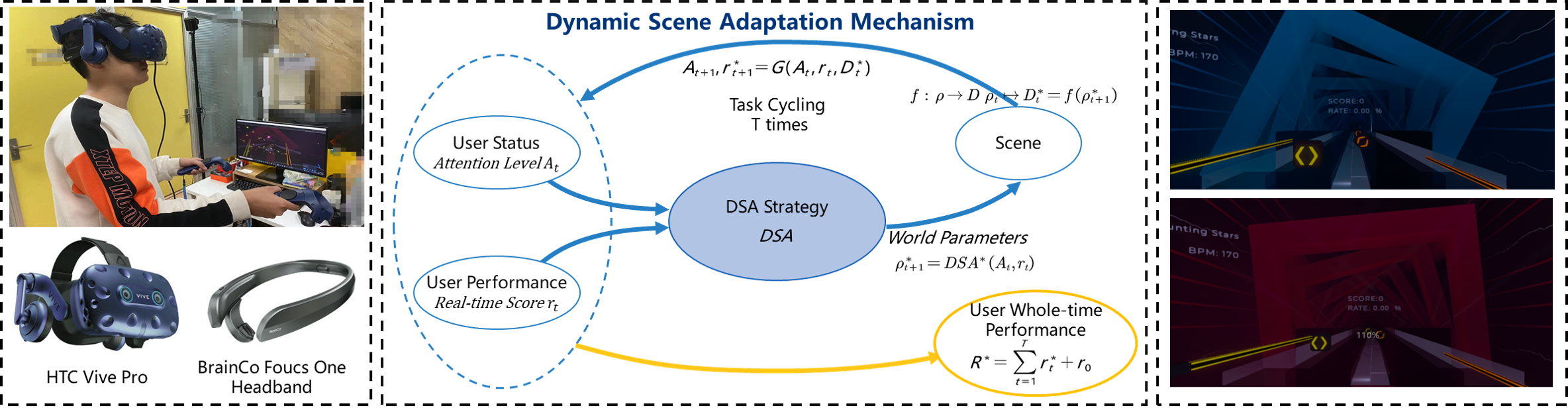}
 \caption{The left part illustrates participants wearing VR and EEG Headband equipment to conduct experiments; the right part is the game scenes before and after DSA strategy adjustment. The middle part is the DSA Mechanism pipeline, which details will be introduced in the second section.}
 \label{fig:DSOsystem}
}

\abstract{
Virtual reality (VR) produces a highly realistic simulated environment with controllable environment variables. This paper proposes a Dynamic Scene Adjustment (DSA) mechanism based on the user interaction status and performance, which aims to adjust the VR experiment variables to improve the user's game engagement. We combined the DSA mechanism with a musical rhythm VR game. The experimental results show that the DSA mechanism can improve the user's game engagement (task performance). 
} 

\CCScatlist{
  \CCScatTwelve{Human-centered computing}{Interaction parad\-igms}{Virtual reality}{};
}

\begin{document}


\firstsection{Introduction}

\maketitle


The environment can have a significant impact on user performance and experience. Virtual reality (VR) produces a highly realistic simulated environment with controllable environment variables.  
Many studies have proved that user status, experience, or performance can be affected by variables such as color, music, and light intensity in the environment~\cite{
liao2020data}.
However, few researchers have tried to dynamically adjust the user's state in real-time by controlling these VR environmental variables.

This work aims to establish a Dynamic Scene Adjustment (DSA) mechanism based on the user interaction state, as illustrated in \autoref{fig:DSOsystem}. 
To enhance the user experience and overall performance (the core of user engagement), the DSA mechanism dynamically modifies the VR environment variables according to the user's instant status and performance.
The inspiration for this research comes from two aspects: a) When we physiologically monitor users' attention in the VR environment, we found that some experiment variables like color significantly impact the user's attention and performance during the task, which has also been reported in previous research~\cite{yoto2007effects,stone1998task}. 
b) Referring to the mature Difficulty Dynamic Adjustment method~\cite{zohaib2018dynamic}, it should also be reasonable to improve the users'  experience and overall performance by dynamically adjusting the VR environment through adaptive user's instant status and performance.

We also designed a DSA game demo based on the open-source VR game Beat Saber\footnote{https://beatsaber.com/}. We invited 13 participants to conduct a total of 18 groups of comparative experiments on the DSA demo game.
The experimental results show that adjusting the color (experiment variable) in the VR environment through adaptive user attention (status) and instant performance can simultaneously improve their experience and overall performance.  
This also proves the feasibility of the DSA mechanism. The user's engagement (experience, status, and overall performance) during the interaction process can be optimized by adjusting the interaction environment parameters.

\section{Dynamic Scene Adjustment Mechanism}
\textbf{Design of the Dynamic Scene Adjustment (DSA) mechanism.}
Users are not only unilaterally affected by the environment; the decisions made at every moment and their status, performance, and experience can also affect the user's state at the next moment. Thus, the DSA mechanism can be described with the following equations.
\begin{equation}
\begin{split} 
& \rho _{t+1}^{*}=\mathrm{DSA}^*\left( A_t,r_t \right) \quad 
f\,\,: \rho \rightarrow D, \rho _t\mapsto D_{t}^{*}=f\left( \rho _{t+1}^{*} \right) 
\\
& A_{t+1},r_{t+1}^{*}=G(A_t,r_t,D_{t}^{*})
\qquad
R^*=\sum_{t=1}^T{r_{t}^{*}+r_0\,\,}
\end{split}
\end{equation}
where $\rho$ is the world parameters, the function $DSA^*\left(\right)$ is the best DSA intervention strategy, $A_{t}$ is users' status, $r_t$ is users' current performance, $r_{t+1}^*$ is the user's performance after enhanced by the DSA mechanism at the next moment, $A_{t+1}$ is the status when the user interacts with the scene adjusted by the DSA mechanism, the function $G\left(\right)$ is the game-play rules. $R^*$ is the best user overall interaction performance, which is also the adjustment goal of the DSA mechanism. $D$ is the user's action, which satisfies the mapping relationship $f$ with $\rho$. The intervention strategy $DSA$ calculates the world parameters with users' status and performance as input. Many intervention strategies exist, and the best strategy $DSA^*$ will obtain the most suitable parameter $\rho^*$. $\rho^*$ will cause the user to take the best action $D^*$. After the user's state and action interact with the gameplay, their new status $A_t$ and performance $r_{t+1}$ will occur. This is the logic of the operation of the DSA mechanism.
Therefore, the DSA mechanism mainly includes three modules.
a) \textbf{World parameters} are the environmental variables that mainly affect the user's performance and experience. 
b) \textbf{Adjustment goal} is the expected state or desired experience of the users (or the anticipated state the examiner hopes participants can complete).
c) \textbf{Intervention strategy} is the calculation function of world parameters, which is the logic about how users achieve the adjustment goal.  

\begin{figure}[tb]
 \centering 
 \includegraphics[width=\columnwidth]{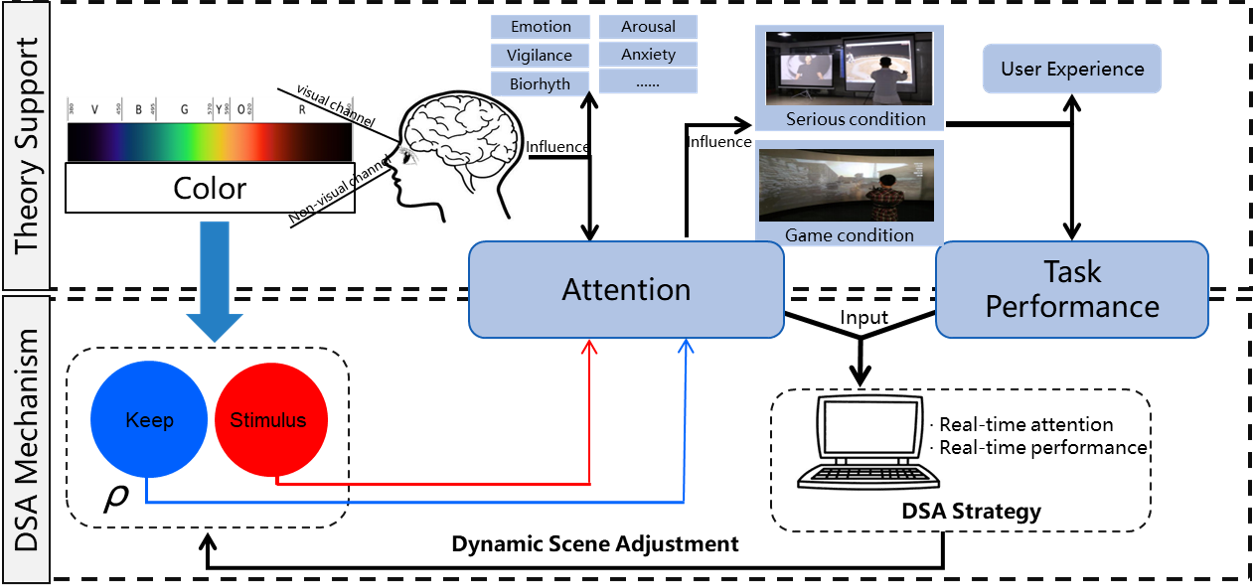}
 \caption{DSA game pipeline, the upper part of the figure is the theoretical basis for adjustment goals and intervention strategies. The lower part is the specific DSA logic applied in the game}
 \label{fig:DSOgame}
\end{figure}

\textbf{Gameplay design with DSA mechanism.}
In this part, we introduce our DSA game to readers according to the three main modules required by DSA, as illustrated in \autoref{fig:DSOgame}.
We selected the background color in the VR environment as the world parameter $\rho$ and set it to either red or blue. Because the selected game is a kind of musical rhythm VR game, it has been proven that color can influence users' attention, \textcolor{red}{Red} can stimulate users' attention, and \textcolor{blue}{Blue} will keep users' attention stable~\cite{yoto2007effects}. Users' attention will influence users' experience~\cite{bian2018exploring} and performance~\cite{wickens2020processing}. We set the users' game score (user performance) in the VR environment as the adjustment goal.
Regarding the intervention strategy, we input the user's real-time attention and instant performance into the DSA strategy module, dynamically adjusting the world parameters selected (color) in this game. Users' instant performance is calculated by the user's score changes within an adjacent time window. The adjustment strategies of the game are illustrated in \autoref{tab:DSOstrategy}. We assume that the user's score ratio in the first time window is $S_t$, while in the second time window is $S_{t+1}$. Users' instant performance $r_{t+1}=S_{t+1}-S_{t}$. The time window size is 2.5 seconds which is the hyper-parameters obtained after decoupling the game.

\section{Experiment Setup and Results analysis}
\textbf{Experiment Setup} The experiment was supervised and guided by the Ethics Committee of UESTC.
13 participants, including 4 female and 9 male students between 20 and 28 years old, completed the comparative experiment. The game with (or without) the DSA mechanism is used as the experimental (control) group. Participants were asked to be involved twice in each experimental and control group experiment in random order.
To avoid the impact of proficiency, all participants must practice several times before the formal experiment. Participants are given enough rest time after each game to avoid the impact of fatigue. The user's attention and task performance (game score) are collected and saved during the experiments. A short user experience interview is conducted after the experiment.

\textbf{Experiment Results}
With the DSA mechanism, 83.3$\%$ of the participants improved their performance. The paired sample t-test results show that the test subjects performed better in the experimental group (M=86.13278, SD=8.79289) than in the control group (M=84.47861, SD=9.7518).
The experimental results show that the \textbf{DSA mechanism can improve the user's task performance in the VR environment.}
In addition, according to interview feedback, almost all participants reported that the experimental group was more excited and they experienced better than the control group.

\section{Conclusion}
In summary, this paper proposes the Dynamic Scene Adjustment (DSA) mechanism for dynamically improved user interaction performance by adjusting the world parameters in the VR environment following the user's real-time status and instant performance.
In addition, we combined the DSA mechanism with a musical rhythm VR game and applied it to VR environment tasks. The experimental conclusions support the rationality of the mechanism. This work can help researchers think about dynamic regulation in VR environments from a new perspective. It can inspire the design of VR therapy, VR education, VR games, and other fields. Future research will expand the experimental data based on the existing DSA mechanism, adding more world parameters. In addition, we will want to establish a universal DSA mechanism through a data-driven approach.

\begin{table}[tb]
  \caption{Dynamic Scene Adjustment Strategy}
  \label{tab:DSOstrategy}
  \scriptsize%
	\centering%
  \begin{tabu}{cccc}
  \toprule
   No. & \makecell[c]{User Performance\\ (Score Ratio)} & \makecell[c]{User Status \\ (Attention Level)} & \makecell[c]{World Parameters \\ (Color)} \\
  \midrule
1 & $S_{t+1}-S_t\geq0$ & High & \textcolor{blue}{Blue}\\
2 & $S_{t+1}-S_t\geq0$ & Low & \textcolor{red}{Red}\\
3 & $S_{t+1}-S_t<0$ & Low & \textcolor{red}{Red}\\
4 & $S_{t+1}-S_t<0$ & High & \makecell[c]{Maintain current \\ scene color }\\
  \bottomrule
\end{tabu}%
\end{table}
%
\acknowledgments{
This work was supported by Development and Application Demon- strations of Digitalized Governance of Local Society for Future Cities Research Program (2021-JB00-00033-GX), and the National Natural Science Foundation of China under Grant NO. 61976156.}

\bibliographystyle{abbrv-doi}

\bibliography{template}

\begin{thebibliography}{1}

\bibitem{bian2018exploring}
Y.~Bian, C.~Yang, C.~Zhou, J.~Liu, W.~Gai, X.~Meng, F.~Tian, and C.~Shen.
\newblock Exploring the weak association between flow experience and
  performance in virtual environments.
\newblock In {\em Proceedings of the 2018 CHI conference on human factors in
  computing systems}, pp. 1--12, 2018.

\bibitem{liao2020data}
H.~Liao, N.~Xie, H.~Li, Y.~Li, J.~Su, F.~Jiang, W.~Huang, and H.~T. Shen.
\newblock Data-driven spatio-temporal analysis via multi-modal zeitgebers and
  cognitive load in vr.
\newblock In {\em 2020 IEEE Conference on Virtual Reality and 3D User
  Interfaces (VR)}, pp. 473--482. IEEE, 2020.

\bibitem{stone1998task}
N.~J. Stone and A.~J. English.
\newblock Task type, posters, and workspace color on mood, satisfaction, and
  performance.
\newblock {\em Journal of Environmental Psychology}, 18(2):175--185, 1998.

\bibitem{wickens2020processing}
C.~D. Wickens.
\newblock Processing resources and attention.
\newblock In {\em Multiple-task performance}, pp. 3--34. CRC Press, 2020.

\bibitem{yoto2007effects}
A.~Yoto, T.~Katsuura, K.~Iwanaga, and Y.~Shimomura.
\newblock Effects of object color stimuli on human brain activities in
  perception and attention referred to eeg alpha band response.
\newblock {\em Journal of physiological anthropology}, 26(3):373--379, 2007.

\bibitem{zohaib2018dynamic}
M.~Zohaib.
\newblock Dynamic difficulty adjustment (dda) in computer games: A review.
\newblock {\em Advances in Human-Computer Interaction}, 2018, 2018.

\end{thebibliography}
\end{document}